\documentclass[fleqn,usenatbib]{mnras}

\usepackage{newtxtext,newtxmath,nicefrac}

\usepackage[T1]{fontenc}

\DeclareRobustCommand{\VAN}[3]{#2}
\let\VANthebibliography\thebibliography
\def\thebibliography{\DeclareRobustCommand{\VAN}[3]{##3}\VANthebibliography}

\usepackage{graphicx}	
\usepackage{amsmath}	

\newcommand{\new}[1]{{{#1}}}

\title[Implications of Eccentric NSBHs]{Astrophysical Implications of Eccentricity in Gravitational Waves from Neutron Star-Black Hole Binaries}

\author[I. Romero-Shaw et al.]{
Isobel Romero-Shaw,$^{1}$\thanks{E-mail: romero-shawi@cardiff.ac.uk}
Jakob Stegmann,$^{2}$
Gonzalo Morras,$^{3}$
Andris Dorozsmai,$^{4}$
and Michael Zevin$^{5,6,7}$
\\
$^{1}$Gravity Exploration Institute, School of Physics and Astronomy, Cardiff University, Cardiff, CF24 3AA, UK\\
$^{2}$DMax Planck Institute for Astrophysics, 
Karl-Schwarzschild-Str.~1, 85748 Garching, Germany\\
$^{3}$Max Planck Institute for Gravitational Physics (Albert Einstein Institute), D-14467 Potsdam, Germany\\
$^{4}$National Astronomical Observatory of Japan, National Institutes of Natural Sciences, 2-21-1 Osawa, Mitaka, Tokyo 181-8588, Japan\\
$^{5}$Adler Planetarium, Department of Astronomy, 1300 S. Lake Shore Drive, Chicago, IL 60605, USA\\
$^{6}$Center for Interdisciplinary Exploration and Research in Astrophysics (CIERA), Northwestern University, 1800 Sherman Ave, Evanston, IL 60201, USA\\
$^{7}$NSF-Simons AI Institute for the Sky (SkAI), 172 E. Chestnut St., Chicago, IL 60611, USA
}

\date{Accepted XXX. Received YYY; in original form ZZZ}

\pubyear{\the\year{}}

\begin{document}
\label{firstpage}
\pagerange{\pageref{firstpage}--\pageref{lastpage}}
\maketitle

\begin{abstract}
The gravitational-wave signal from the neutron star-black hole (NSBH) merger GW200105 is consistent with this binary having significant orbital eccentricity close to merger.
This raises the question of how eccentric NSBHs form. 
Compact binaries that \new{evolve in isolation} radiate away any orbital eccentricity long before their gravitational-wave signal enters the sensitive frequency range of the LIGO-Virgo-KAGRA detector network. 
Meanwhile, dynamical environments---which can be conducive to mergers on eccentric orbits---produce very few NSBHs.
Here, we focus on a formation channel that efficiently produces NSBHs with both misaligned spins and significant eccentricity close to merger: isolated field triples.
We estimate the minimum measurable eccentricity of NSBHs at $10$~Hz orbit-averaged gravitational-wave frequency, $e_{\mathrm{min},10}$, finding that for GW200105, GW200115, and GW230529-like systems, $e_{\mathrm{min},10}$ is $\mathcal{O}(0.01)$. For a GW190814-like unequal-mass binary with significant higher-order mode content, $e_{\mathrm{min},10} = 0.003$; this is an order of magnitude lower than when higher modes are not present.
For dominant-mode signals from binaries with $m_2=1.5$~M$_\odot$ and total masses from $3\,{\rm M_\odot} \leq M \leq50\,\rm M_\odot$, we find $0.008 \leq e_{\mathrm{min},10} \leq 0.022$. The relationship between $M$ and $e_{\mathrm{min},10}$ is linear when the binaries are non-spinning. When the binaries are maximally spin-precessing, $e_{\mathrm{min},10}$ decreases as mass ratio becomes more unequal.
We estimate the sensitivity of a quasi-circular \new{aligned-spin} templated search to NSBH mergers from field triples, finding that we recover only \new{$46\%$ of systems that would have been detected with a search containing the full physics of the injected population.}
Finally, we show that if $\sim\nicefrac{1}{3}$ of present NSBH detections are measurably eccentric, then \new{$\geq40\%$} are consistent with an isolated field triple origin.
\end{abstract}

\begin{keywords}
transients: black hole - neutron star mergers -- gravitational waves -- stars: neutron -- stars: black holes
\end{keywords}



\section{Introduction}


The LIGO-Virgo-KAGRA (LVK) Collaboration \citep{LVK} has reported the confident detection of three neutron star-black hole (NSBH) mergers via their gravitational-wave (GW) signals. GW200105 and GW200115 \citep{NSBHs}, reported during the LVK's third observing run (O3), have primary component masses consistent with being black holes (BHs), and secondary masses consistent with being neutron stars (NSs). GW230529 \citep{GW230529}, reported during the fourth observing run (O4), has an NS secondary and a primary with an ambiguous mass of $2.5 \lesssim m_1 \lesssim 4.5$~M$_\odot$, which resides within the ``lower mass gap'' between NS masses and BH masses; GW230529 is thought to most likely be a low-mass NSBH, given the theoretical Tolman-Oppenheimer-Volkoff (TOV) upper limit to NS masses ($M_\mathrm{TOV} = 2.25^{+0.08}_{-0.07} \rm M_\odot$ \citep{2024PhRvD.109d3052F}; see also \citep{1939PhRv...55..364T, PhysRev.55.374, 1996ApJ...470L..61K, 2018ApJ...852L..25R}). Other candidates are consistent with being of NSBH origin, although with lower probability. GW190814 \citep{GW190814}, detected in O3, is an ambiguous merger with a secondary residing in the lower mass gap; GW190814, again due to the TOV mass limit, is thought most likely to be a binary BH merger. There are six lower-significance or ambiguous NSBH candidates in the second observing run (O2) and O3 \citep{GWTC-2, GWTC-2-1, GWTC-3}. In the engineering run preceding O4, the NSBH candidate GW230518\_125908 was detected \citep{GWTC-4}.

Renewed interest in one of these NSBHs, GW200105, has recently arisen due to follow-up studies finding evidence for measurable orbital eccentricity in this GW signal \citep{Fei2024,2025arXiv250315393M, planas2025, Kacanja2025, 2025arXiv250812460J, 2025arXiv250926152T}. The Bayes factor in favour of the eccentric hypothesis over the quasi-circular hypothesis is low: for example, \citet{planas2025} find $\log_{10}$ Bayes factors from $\approx0.1$ to $1.2$ depending on prior settings, indicating negligible to mild preference of the eccentric hypothesis. Nonetheless, these findings motivate further investigation into the formation, detection, and identification of eccentric NSBHs. 

In general, an NSBH merger may form as a result of the isolated evolution of a massive stellar binary \citep[e.g.,][]{1998A&A...332..173P, 2002ApJ...572..407B, 2012ApJ...759...52D, 2021ApJ...920L..13B, 2024A&A...683A.144X, 2022MNRAS.513.5780C}, in which case the binary would merge with undetectable eccentricity \citep{Peters1964, 2002ApJ...572..407B}. Alternatively, the binary may form via a pathway that allows it to retain significant eccentricity close to merger. For binary BHs, detectable eccentricities at a GW frequency of $10$~Hz, $e_{10}$, are possible when formation and merger occurs in densely-populated environments like globular or nuclear star clusters and active galactic nuclei (AGN) \citep[e.g.,][]{2018PhRvL.120o1101R, 2019ApJ...871...91Z, 2023:Chattopadhyay:NSCs, Tagawa2021_ecc}. However, NSBH mergers from globular cluster environments are not expected to contribute significantly to LVK observations: the predicted local merger rate densities $\lesssim 0.17$~Gpc$^3$~yr$^{-1}$\citep{2013MNRAS.428.3618C, Bao-Minh2020, 2020ApJ...888L..10Y} are low compared to the NSBH merger rate inferred from GW observations, $9.1$-$84 $~Gpc$^3$~yr$^{-1}$\citep{GWTC-4-astro-pop}. Moreover, the few NSBH binaries that could be produced in these environments do not retain measurable eccentricities close to merger \citep{ArcaSedda2020}. NSBHs can be formed efficiently in young star clusters, with rates $\sim 28$~Gpc$^3$~yr$^{-1}$, but are ejected and merge in the field, leading them to have negligible eccentricities at merger \citep{2020MNRAS.497.1563R}. Detectably-eccentric NSBHs may be produced in nuclear clusters, but, similarly, local merger rate densities are predicted to be very low, $\lesssim 0.10$~Gpc$^3$~yr$^{-1}$ \citep{2019MNRAS.488...47F}. Rates of NSBH mergers in AGN are potentially high and consistent with observed LVK rates, but their eccentricities are uncertain \citep{2020MNRAS.498.4088M}.

An NSBH that is observed with detectable close-to-merger eccentricity may instead result from the evolution of isolated hierarchical triples in the galactic field, where the tertiary companion can drive large-amplitude von~Zeipel-Kozai-Lidov (ZKL) oscillations of the NSBH eccentricity in the inner binary \citep{Zeipel1910,Kozai1962,Lidov1962,Naoz2016}. As a result, efficient gravitational-wave emission at highly eccentric pericentre passages shrinks the orbit of the NSBH, effectively decouples it from further perturbation from the tertiary, and can lead to an inspiral and merger with residual eccentricity. Owing to the large abundance of massive stellar triples in the field \citep{Moe2017,Offner2023}, previous studies find that this process can efficiently produce NSBH mergers at rates consistent with or higher than that inferred from LVK observations \citep{FragioneLoeb:2019:triples, HamersThompson:2019:triples,StegmannKlencki2025}. Thus, residual eccentricities $>10^{-4}$ at $10$~Hz, and spins that are misaligned with the orbital angular momentum of the binary, are characteristic of NSBHs formed through the field evolution of triple stars \citep{StegmannKlencki2025}.

Current LVK GW searches are inefficient for detecting inspiral-dominated binaries with significant eccentricities, with the loss of sensitivity worsening for higher eccentricities and lower masses \citep{2010PhRvD..81b4007B, 2024:Divyajyoti:blind-spots, 2024PhRvD.110d4013G, Bhaumik:2025:cWBsearchsensitivity}. LVK search methods are either \textit{modelled}, using quasi-circular aligned-spin waveform templates \citep[e.g.,][]{MBTA, PyCBC, GSTLAL, SPIIR}, or \textit{unmodelled}, looking for correlated excess power in the data \citep{2016PhRvD..93d2004K, 2021SoftX..1400678D}. 

Highly-eccentric merger-dominated signals with minimal inspiral may be detected in a burst search \citep[see, e.g.,][and references therein]{2025arXiv250712374L}, while low-eccentricity merger-dominated signals may be picked up in a quasi-circular modelled search due to the close match with a quasi-circular aligned-spin template in the merger. Meanwhile, modelled searches are less sensitive to lower- and more unequal-mass binaries if eccentricity is excluded from search templates; such signals are also less likely to be flagged during a burst search due to the relatively lower-powered merger \citep[e.g.,][]{2010PhRvD..81b4007B, 2013PhRvD..87l7501H, 2024:Divyajyoti:blind-spots}. Even when both burst and modelled searches are used, one loses a significant fraction of eccentric sources \citep{2024PhRvD.110d4013G}. Thus, if there are many NSBH mergers with non-negligible orbital eccentricity---as predicted from field triple evolution---then there is a high risk of not detecting them in LVK searches. Since injection campaigns used to estimate selection effects also do not include eccentricity, the NSBH merger rate inferred from GW observations may thus be artificially skewed to lower values due to the loss of search sensitivity to eccentric NSBHs.

Several modelled searches targeting inspiral-dominated \new{(i.e., long duration) eccentric signals} have been developed \citep[e.g.,][]{2020ApJ...890....1N, 2021ApJ...915...54N, 2025PhRvD.111j3018D, 2025arXiv250805018W}. \citet{2025PhRvD.111j3018D} specifically target aligned-spin eccentric NS-containing binaries, requiring the signal to trigger a detection in more than one instrument. 
Their template bank covers $e_{22,10} \leq 0.46$, where the $22,10$ subscript indicates that the eccentricity is measured when the dominant $l=m=2$ ($22$) mode GW frequency reaches $10$~Hz. Since GW200105 was only detected in LIGO Livingston (although Virgo data was also analysed), it was not recovered by this search. This search is also restricted to detector-frame total masses of $\leq 10$~M$_\odot$, which, in combination with the eccentricity constraint, puts $\gtrsim80\%$ of the NSBH masses and eccentricities predicted by \citet{StegmannKlencki2025} outside of the searched range even before considering the loss of search sensitivity due to the misaligned spins expected from triples. In an aligned-spin, inspiral-only targeted search for eccentric NSBHs in O3, \citet{Phukon:2025cky} recover GW200105 at higher significance than in an equivalent quasi-circular search, with no other significant candidates. However, this search restricts to detector-frame primary masses $5 \leq m_1 \leq 15$~M$_\odot$ and eccentricities $e_{22,20} \leq 0.2$ (corresponding to $e_{22,10}\lesssim 0.35$), thereby excluding $\gtrsim72\%$ of the parameter space predicted by \citet{StegmannKlencki2025} before considering misaligned spins. Precessing-spin searches have also been developed specifically for NSBHs, revealing that current LVK search methods may be artificially biased against detecting this flavour of binary due to neglecting misaligned spins \citep{2025arXiv250309773H}.

Long inspirals are both a blessing and a curse when it comes to detecting and measuring eccentricity. One the one hand, eccentric inspirals are less likely to be \textit{detected} by quasi-circular searches, since the mismatch between an eccentric signal and a circular template leads to a reduction in SNR and astrophysical probability. On the other hand, precisely because of this larger overall difference to a quasi-circular signal, they increase our ability to \textit{measure} eccentricity if analysed using an eccentric waveform model: the eccentricity of the signal becomes easier to measure if more cycles are in-band \citep[e.g.,][]{2020CQGra..37v5015M, 2020MNRAS.496L..64R, 2023:Romero-Shaw:Ecc-or-precc}. Furthermore, systems observed earlier in their evolution are more likely to retain measurable eccentricity, as they have undergone less GW-driven circularization. The source property that has the dominant influence on signal duration is the mass: lower-mass systems spend longer in-band because they merge at higher frequencies. Additionally, systems with more unequal masses have more cycles in-band. Accordingly, lower-~and more unequal-mass binaries, such as NSBHs, can also be expected to have lower minimum detectable eccentricities.

Similarly, GW signals that contain higher-order modes (HMs) may have reduced detectability due to searches using only dominant-mode templates, but also increase the information extractable from the signal. HMs are most prominent in GW signals from unequal-mass binaries, making unequal-mass NSBHs possible HM candidates. While none of the three confident NSBH mergers detected so far show signs of strong HM content, one possible NSBH candidate---GW190814---does contain HMs \citep{GW190814, 2023PhRvD.107h4026G}. GW190814 originated from an unequal-mass compact binary merger, with source-frame masses $m_1\approx23$~M$_\odot$ and $m_2\approx2.6$~M$_\odot$. The secondary mass of GW190814 lies in the ``lower mass gap'', and its identity---NS or BH---is ambiguous. In GW signals from eccentric binaries, each mode present in the signal acquires overtones. In other words, signals with HMs contain more information about the eccentricity of the binary. As we demonstrate, this can enable the measurement of lower eccentricities than is possible when only the dominant mode is considered.


In this paper, we estimate the minimum detectable eccentricity for NSBH-like mergers. For nonspinning primaries, this correlates with total mass, and tends to be $\mathcal{O}(0.01)$ for NSBH-like binaries with network matched-filter SNRs of $\rho\approx24$. If the binary exhibits significant spin-induced precession in addition to eccentricity, the measurability of eccentricity improves for higher-mass binaries. This is because binaries with more unequal mass ratios have more spin-induced precession cycles in-band, which are altered by the presence of eccentricity. If higher-order modes are present in the signal, the minimum detectable eccentricity can be $\approx1$ order of magnitude lower. We assess the detectability of a population of NSBH mergers produced in field triples, finding that the majority are not detected due to their relatively high eccentricities. Assuming a quasi-circular modelled search only, we \new{detect only $46\%$ of the signals that could have been recovered using a search that contains the full physics of the population}. By modelling the detection of NSBH mergers as a Poisson process, we find that if GW200105 was formed in an isolated field triple, then all NSBHs observed so far may have evolved via the same mechanism. 

We introduce our methodology in Section \ref{sec:methods}, describe our results and  their astrophysical implications in Section \ref{sec:results}, and conclude in Section \ref{sec:discussion}. 

\section{Methods}
\label{sec:methods}

For both detectability and measurability studies presented here, we use the gravitational waveform model \textsc{pyEFPE} \citep{pyEFPE}, an inspiral-only approximant containing the effects of eccentricity, a variable mean anomaly parameter, and misaligned spins. 
Studies with \textsc{pyEFPE} were the first to demonstrate evidence for eccentricity in NSBH signal GW200105 \citep{2025arXiv250315393M}.
\textsc{pyEFPE} is computationally efficient, making it appropriate for population recovery studies requiring $\mathcal{O}(10^4)$ or more waveforms to be generated.
While \textsc{pyEFPE} is limited by being an inspiral-only waveform model without HMs, merging NSBHs have inspiral-dominated signals, and confident NSBH signals so far have not demonstrated evidence for HMs.
To assess the impact of including higher-order modes on the measurability of eccentricity, we use \textsc{SEOBNRv5EHM} \citep{SEOBNRv5EHM,SEOBNRv5EHM:theory}, an effective one-body full inspiral-merger-ringdown waveform model with eccentricity and aligned spins, that has also been used to demonstrate evidence for eccentricity in GW200105 \citep{Kacanja:2025:NSeventseccentric,2025arXiv250812460J}. 

We generate waveforms with minimum and reference frequencies of $10$~Hz, and inject them into simulated O4-sensitivity LVK detector data using the Bayesian inference library \textsc{Bilby} \citep{bilby, bilby2}. 
When generating waveforms, we pass in eccentricity defined at an orbit-averaged frequency of $f_{22} = 10$~Hz.

\subsection{Measurability of eccentricity in NSBHs}

\begin{table*}
    \centering
    \begin{tabular}{c|c|c|c|c|c}
         Variable & fixed $m_2$ & GW190814-like & GW200105-like & GW200115-like & GW230529-like \\
         \hline
         $M$ [M$_\odot$] & variable & 25.78 & 13.00 & 7.30 & 5.00 \\
         $m_1$ [M$_\odot$] & variable & 23.2 & 11.5 & 5.8 & 3.6 \\
         $m_2$ [M$_\odot$] & 1.5 & 2.58 & 1.5 & 1.5 & 1.4 \\
         $q$ & variable & 0.099 & 0.115 & 0.233 & 0.280 \\
         $\chi_1$ & 0.0 & 0.07 & 0.06 & -0.31 & 0.0 \\
         $\chi_\mathrm{p}$ & variable & 0.0 & 0.0 & 0.0 & 0.0 \\
         $\theta_{JN}$ &  0.0 &  0.8 & 2.36 & 0.76 & 2.36 \\
         $d_\mathrm{L}$ [Mpc] & variable & 249 & 296 & 300 & 201 \\
         $f_\mathrm{samp}$ [Hz / $2048$] & 6 & 2 & 2 & 4 & 6 \\
         $D$ [s] & 128 & 16 & 32 & 64 & 128 \\
    \end{tabular}
    \caption{Parameter settings in the eccentricity measurability studies we report here. Eccentricity $e_{22,10}$ is variable in all studies, so is not included in this table. Parameters that are shared between all studies---dimensionless secondary spin ($\chi_2=0.0$), right ascension ($\alpha=1.02$), declination ($\delta=5.73$), geocent time ($t_0 = D-2$~s), phase and polarisation ($\phi=\psi=0$)---are not included in this table. Spins are assumed to be aligned with the orbital angular momentum for event-like injections. For fixed $m_2$ studies we inject both zero-spin signals and signals with a primary that is maximally spinning and misaligned.}
    \label{tab:settings}
\end{table*}

We follow \citet{Lower2018} to estimate the threshold at which eccentricity becomes detectable:

\begin{equation}
    1 - \mathcal{O} \gtrsim \rho_0^{-2} ,
\end{equation}

\noindent where $\rho_0$ is the optimal matched-filter SNR of a signal with identical source properties but zero eccentricity, and $\mathcal{O}$ is the overlap between this quasi-circular waveform $h_0$ and a waveform $h_e$ with eccentricity, maximising over reference phase ($\phi$) and geocentric time ($t_0$):

\begin{equation}
    \mathcal{O} = \max_{t_0, \phi} \frac{\langle h_0 | h_e \rangle}{\sqrt{\langle h_0 | h_0 \rangle\langle h_e | h_e \rangle}}.
\end{equation}

We vary the sampling frequency $f_\mathrm{samp}$, maximum frequency $f_\mathrm{max}$, and data duration $D$ depending on the specific systems under investigation. Note that for \textsc{pyEFPE} the maximum frequency of the waveform is automatically set to the GW frequency of innermost stable circular orbit (ISCO), and for \textsc{SEOBNRv5EHM} we set $f_\mathrm{max} = \frac{1}{2}f_\mathrm{samp}$. 

Waveform and parameter settings are given in Table \ref{tab:settings}. For fixed $m_2$ studies, we investigate a range of values of $q$ between $1/30$ and $1$. We also alter the luminosity distance such that the SNR of the quasi-circular signal is relatively consistent, with $\rho_{0, \mathrm{H1}} \approx 22$. We vary eccentricity on a log-uniform grid in the range $10^{-6} \leq e_{22,10} \leq 0.4$, and for each eccentricity point, we marginalise over the anomaly parameter. We inject into a three-detector network of LIGO Livingston, LIGO Hanford and Virgo with O4 sensitivity.

When using \textsc{SEOBNRv5EHM} with the settings described above, we observe oscillations in the overlap that, inconveniently, often coincide with the sensitivity threshold. We find that the central value around which the overlap oscillates is sensitive to both sampling frequency and inclination, with higher sampling frequencies and more face-on orientations having lower values, indicating that they have a better match with their quasi-circular counterparts. The sensitivity to $\theta_\mathrm{JN}$ is explained by the presence of higher-order modes in this waveform, and we believe the sensitivity to $f_\mathrm{samp}$ is a result of small changes to the waveform initial conditions when varying this parameter. 

To demonstrate the influence of these effects, we compare the overlaps obtained using $f_\mathrm{samp} = 2048$~Hz vs. $f_\mathrm{samp} = 2048 \times 10$~Hz for a GW190814-like injection in the left panel of Figure \ref{fig:compare-oscillations} in the Appendix. While for the GW190814-like injection this increase in $f_\mathrm{samp}$ causes lower-amplitude overlap oscillations at lower $e_{22,10}$, the problem persists for the other injections. In the right-hand panel, we compare the overlaps obtained for a GW200105-like injection with $\theta_\mathrm{JN}=2.36$ and $\theta_\mathrm{JN}=0$. Regardless of sampling frequency or inclination, we find that the value of $e_{22,10}$ above which there are no more instances of $1 - \mathcal{O} < \rho_0^{-2}$ remains similar. Therefore, we use parameters as given in Table \ref{tab:settings} and report the minimum measurable eccentricity as the $e_{22,10}$ above which $1 - \mathcal{O}$ is strictly $> \rho^{-2}_0$.

\subsection{Detectability of eccentric NSBHs}

\begin{figure*}
    \centering
    \hspace{-30pt}
    \includegraphics[width=0.9\linewidth]{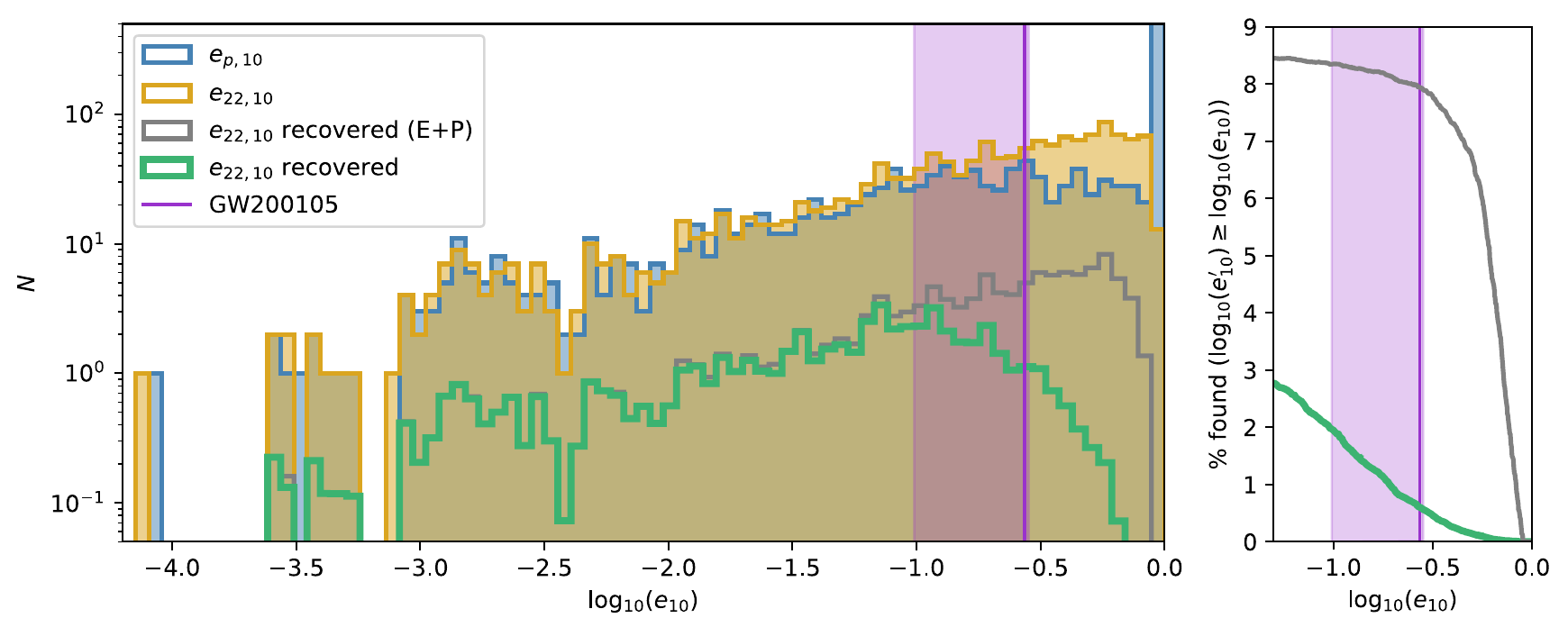}
    \caption{\textit{Left:} The translucent blue filled histogram shows the eccentricity defined at a peak GW frequency of $10$~Hz, output by the triple simulations of \citet{StegmannKlencki2025}. The translucent orange filled histogram shows these same eccentricities but evolved to the $10$~Hz $22$-mode frequency using the standard eccentricity definition scripts of \citet{Vijaykumar2024} (see also \citet{2023:Shaikh:ecc}) or the analytic solutions of \citet{2021PhRvD.104j4023T}, assuming no redshifting. The green unfilled histogram shows the source-frame eccentricity of systems that are recovered by the optimal quasi-circular aligned-spin search described in the text, averaging every injected system over $500$ possible $d_L$---$\ell$ pairs and $10^7$ sets of extrinsic parameters, taking into account the effects of redshifting in the injection. \new{The grey histogram shows the results of the same search, but this time including both eccentricity and spin-precession (E+P) in the template waveforms.} Our mock quasi-circular search recovers \new{$3.9\%$ of the simulated population, representing $46\%$ of signals recovered in the ``E+P'' search, which recovers $8.5\%$}. The vertical purple line shows the median $e_{22,10}$ recovered in \citet{2025arXiv250315393M} for GW200105 using the same waveform model as is used here, and the translucent purple band spans the $90\%$ credible interval around this median. \textit{Right:} The found percentage of events with $\log_{10}(e'_{10})$ equal to or greater than the value shown on the horizontal axis in the quasi-circular search (green) and eccentric spin-precessing search (grey).}
    \label{fig:e_10_recovered_pyefpe}
\end{figure*}

We compute the detectability of eccentric NSBHs using one O4-sensitivity LIGO Livingston detector as follows.

We take the parameters of the merging NSBHs from the massive triple star simulations of \citet{StegmannKlencki2025}; interested readers should consult that paper for full details of the simulations. The $N_{\rm sim}=1542$ NSBH mergers in these simulations are representative results from the evolution of a wide range of isolated triple star systems, because only a narrow subset of such systems produces a merging NSBH binary.

Each simulated system comprises three stars: two in an inner binary with masses $m_{*,1} \geq 10$~M$_\odot$, $m_{*,2} \geq 5$~M$_\odot$, and one less massive tertiary with $m_{*,3} < 5$~M$_\odot$. The mass of the primary is sampled from a \citet{Kroupa2001} mass function in the range $10 \leq m_1 \leq 100$~M$_\odot$. There is no correlation between the black hole mass $m_1$ and the final eccentricity at $10$~Hz.  When the secondary becomes an NS, its mass is fixed to $m_2=1.5$~M$_\odot$. The simulations are carried out at a fixed metallicity $Z\approx0.1Z_\odot$: for isolated NSBH mergers, metallicity has a weak impact on merger rates and mass distributions \citep[e.g.,][]{2019MNRAS.490.3740N, 2021ApJ...912L..23R, 2022MNRAS.516.5737B, 2023MNRAS.524..426I}, and it is assumed here that the same is true for triples. 


The inner binary evolution proceeds in a two-stage process typical of NSBH formation \citep{2021ApJ...920L..13B, 2024A&A...683A.144X}: (i) the primary becomes a BH while the secondary is still on the main sequence, then (ii) the OB secondary becomes an NS. The birth of the NS---the most disruptive step in the evolution of these systems---acts to filter out triples that are not closely bound. There are two main causes of disruption: first, the inner binary could be disrupted due to the natal kick of the newborn NS; second, the recoil kick to the centre-of-mass of the nascent NSBH due to the mass loss of the secondary could detach the binary from the tertiary, with a kick magnitude roughly inversely proportional to the inner binary separation $a_1$ \citep[cf.][]{LuNaoz2019}. \citet{StegmannKlencki2025} find 
that the natal kick imposes an strict upper limit on the inner orbital separation ($a_1\lesssim10^2\,\rm AU$), while only weakly restricting the outer orbit ($a_2\lesssim10^4\,\rm AU$) as a result of the unequal mass ratio of the newly born NSBH. However, the distribution of surviving NSBH triples is strongly skewed by the requirements imposed by the recoil kick to the centre-of-mass: the tighter the inner binary, the closer the tertiary must be in order for the system to remain bound as a triple. For the surviving systems, $a_2/a_1$, peaks at $\approx10$ with a sharp cut-off at $10^3$.

As a result, the surviving NSBHs are confined in closely-bound, compact triples. The strength of the octupole term in the secular equations of motion for hierarchical triples scales with both $a_1/a_2$ and $(1-q)/(1+q)$ \citep{2013MNRAS.431.2155N}. Consequently, due to the generally more unequal mass ratios of NSBHs, octupole terms in the evolution become more important \citep{Liu2015, Naoz2016, StegmannKlencki2025}. This makes the occurrence of ZKL oscillations less conditional on the outer binary plane being close-to-perpendicular to the tertiary orbital plane, thereby giving rise to higher eccentricities from a wide range of initial conditions \citep{StegmannKlencki2025}. 


Using scripts from \citet{Vijaykumar2024}, we convert the simulation-output eccentricity defined at the \citet{Wen2003} peak GW frequency of $10$~Hz, $e_{p,10}$, to an orbit-averaged 22-mode ($l=m=2$) frequency \citep{2023:Shaikh:ecc} of $10$~Hz, $e_{22,10}$.
The former frequency definition corresponds to that of the harmonic at which the maximum power is emitted for an eccentric binary, and can be thought of as the GW frequency at periastron \citep[e.g.,][]{Wen2003, Hamers2021, Vijaykumar2024}. 
It is recommended that the latter definition is used instead for GW observations: this increases monotonically over the course of the inspiral, avoids the low- and high-eccentricity breakdowns of the \citet{Wen2003} definition, and enables eccentricity to be measured directly from the gravitational waveform \citep{2023:Shaikh:ecc}.

Of the 1542 binaries, 70 ($4.5\%$ of the total) cause the conversion script to fail due to numerical issues in those scripts, which occur for systems with very  small semi-latus rectum $p = a(1 - e^2)$. 
For most of these (53 of 70), we are able to use the formulae of \citet{2021PhRvD.104j4023T} to evolve to the converted 22-mode eccentricity; for the remaining 17, we discard the systems. 
The difference between the two definitions can be seen between the blue and orange histograms in the left-hand panel of Figure \ref{fig:e_10_recovered_pyefpe}. Discarded systems are included in the histogram of $e_{p,10}$, but not in the histogram of $e_{22,10}$.

We first wish to represent a realistically-distributed population of triple-origin NSBH mergers.
For each of the $N_\mathrm{sim}$ binaries in the simulation output, we generate $N_\mathrm{pop}=500$ pairs of luminosity distance $d_\mathrm{L}$ and relativistic anomaly $l$ values.
We assume a uniform distribution between $0$ and $2\pi$ for $l$. 
\new{We use the \texttt{UniformSourceFrame} prior from \textsc{bilby} to generate our $d_L$ distribution in the range $1$ to $1000$~Mpc; this corresponds to a redshift distribution that is uniform in comoving volume and source-frame time. We set this relatively low upper limit on $d_L$ to mitigate computational costs; the maximum distance to which the heaviest system output by the simulation can be detected, if we assume $e_{22,10}=0$, is just below $1000$~Mpc (at $1000$~Mpc, this system has a matched-filter SNR of $\simeq6$).} 

We then build on the method used in \citet{Zevin2021}\new{, which is itself built on the prescriptions in \citet{1993PhRvD..47.2198F},} to compute the detectability of our simulated population. 

\new{To start, we calculate the maximum SNR possible for each system, given quasi-circular aligned-spin recovery templates; we will temper this with realistic extrinsic parameter distributions later.} For each $i$ of the $N_\mathrm{pop}$ pairs, we inject the eccentric NSBH signal with the optimal source orientation of face-on, directly overhead, and \new{circularly} polarised.
We compute the maximum matched-filter signal-to-noise ratio (SNR), $\rho_{\mathrm{max}, i}$, obtained with the quasi-circular waveform template.
We set the sampling frequency to $4096$~Hz, and allow $128$~s of data when generating waveforms.

While \citet{Zevin2021} computed the SNR against a quasi-circular waveform with all other parameters matching, the SNRs of inspiral-dominated NSBH signals are far more sensitive to small changes between the matched templates. 
Before computing the \new{maximum possible} SNR, we therefore minimise the mismatch between the eccentric signal and quasi-circular template by varying the geocent time $t_0$, phase $\phi$, polarisation $\psi$, and amplitude $\mathcal{A}$ \new{following the method described in \citet{2016PhRvD..94b4012H}}.
We also set the $x$ and $y$ components of the primary's spin vector to $0$ in the quasi-circular template.
Other parameters of the quasi-circular aligned-spin template waveform match those of the eccentric and arbitrarily spinning injection.
Attempts to minimise the mismatch over mass and spin parameters yielded negligible improvements over just optimising $(t_0, \phi, \psi, \mathcal{A})$, so were left out of final results.
Our maximum matched-filter SNRs may still be lower than the true possible maximum, since we do not optimise over the whole range of intrinsic parameters.
The detection efficiency we estimate here should therefore be taken as a lower limit.

\new{To continue with more realistic SNR estimates,} we then draw $N_\mathrm{ext}=10^7$ realisations of extrinsic parameters other than $d_\mathrm{L}$ and $l$ from the default \textsc{bilby} prior distributions for binary BHs \citep{bilby2}.
We compute the detector projection factor $\Theta$ for each $j$ of these realisations \citep{1993PhRvD..47.2198F}:

\begin{equation}
    \Theta_j^2 = 4 \left[ F_{+,j}^2(1+\cos^2{\iota_j})^2 + 4 F_{\times,j}^2 \cos^2{\iota_j} \right],
\end{equation}

\noindent where the explicit dependence of the detector response functions $F_+$ and $F_\times$ on extrinsic parameters other than $\iota$ has been dropped from the notation for brevity. 

Setting an SNR threshold $\rho_\mathrm{thresh} = 8$, we compute the detection probability $P_{\mathrm{det}, i}$ for each pair:

\begin{equation}
    P_{\mathrm{det}, i} = \sum_{j=1}^{N_\mathrm{ext}} \mathcal{H}\left[\frac{\Theta_j}{4}\rho_{\mathrm{max},i} - \rho_\mathrm{thresh}\right],
\end{equation}

\noindent where $\mathcal{H}$ is the Heaviside step function.

For each $k$ of the $N_\mathrm{sim}$ mergers output by the simulation, we assign the final weight

\begin{equation}
    w_k = \frac{1}{N_\mathrm{pop}}\sum_{i=1}^{N_\mathrm{pop}} P_{\mathrm{det}, i}.
\end{equation}


\noindent This weight describes the normalised probability that event $k$ is detected, marginalised over our assumed distributions of extrinsic parameters.

Note that we assess the detectability of signals from NSBH sources based only on their matched-filter SNR, which is the optimal ranking statistic in Gaussian noise.
In reality, templated searches must contend with non-Gaussian transients (``glitches'') that can produce spuriously high SNRs. To mitigate the impact of this, searches evaluate the consistency of the data with the template using methods that look beyond the raw SNR; for example, by checking whether the SNR accumulates with frequency as predicted by the waveform model \citep[e.g.,][]{Allen:2004gu, 2012PhRvD..85l2006A}. 
Since the frequency evolution of eccentric signals deviate from quasi-circular templates, these consistency checks tend to down-weight their triggers even if the SNR is high.
Because our analysis does not account for this signal–template inconsistency penalty, our detectability estimates for a triple-origin eccentric NSBH population may be over-optimistic.

\section{Results}
\label{sec:results}

\subsection{Minimum measurable eccentricities for NSBHs}

\begin{figure}
    \centering
    \includegraphics[width=0.9\linewidth]{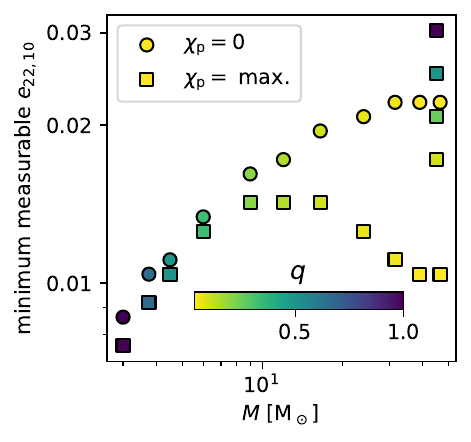}
    \caption{Scatter plot of total mass vs estimated minimum measurable eccentricity calculated via the method described in the text, for injections of \textsc{pyEFPE} waveforms with either minimum or maximum values of $\chi_\mathrm{p}$~\citep{Schmidt:2014iyl}. The H1 SNR of the quasi-circular versions of these injections is within $\pm 0.4$ of $\rho_{0, H1}=22$. For a fixed SNR, the minimum detectable eccentricity correlates with total mass when $\chi_\mathrm{p}=0$. When $\chi_\mathrm{p}$ is maximal ($\chi_1^\perp = 0.99$; we assume the secondary is non-spinning), the relationship is not linear, and measurability improves for more extreme mass ratios. The colour range indicates the mass ratio $q$.}
    \label{fig:scatter-plot-pyefpe}
\end{figure}

\begin{table}
    \centering
    \begin{tabular}{c|c|c}
          & \texttt{pyEFPE} & \texttt{SEOBNRv5EHM} \\
         \hline
         GW190814-like & 0.023 & 0.0032 \\
         GW200105-like & 0.034 & 0.026 \\
         GW200115-like & 0.030 & 0.029 \\
         GW230529-like & 0.020 & 0.023 \\
    \end{tabular}
    \caption{Overlap estimate of minimum measurable eccentricity measured at $10$~Hz orbit-averaged frequency for injected waveforms with parameters similar to NSBH candidates detected by the LVK.}
    \label{tab:minimum-detectable-ecc-nsbhs}
\end{table}

In Table \ref{tab:minimum-detectable-ecc-nsbhs}, we give the estimated minimum measurable eccentricity inferred for the event-like injections we perform, computed with both waveform models. For NSBH-like injections, the minimum detectable eccentricity computed with \texttt{SEOBNRv5EHM} is very similar to that computed with \texttt{pyEFPE}. However, for the GW190814-like injection, which is known to have significant HM content, the estimate obtained with \texttt{SEOBNRv5EHM} is an order of magnitude lower than that obtained with \texttt{pyEFPE}. We confirm by running a 22-mode-only injection of a GW190814-like waveform with \texttt{SEOBNRv5EHM} that the improvement is due to the inclusion of HMs: without them, the minimum detectable eccentricity is $e_{22,10}=0.018$.

We provide a scatter plot of minimum measurable eccentricities obtained with \texttt{pyEFPE} for systems with $m_2=1.5$~M$_\odot$ and varying total mass $M$ in Figure \ref{fig:scatter-plot-pyefpe}. When $\chi_\mathrm{p}=0$, the relationship between total mass and minimum measurable eccentricity is straightforward: smaller total mass gives a better estimated sensitivity to eccentricity for a fixed SNR. To test how this varies with spin-induced precession, we maximise $\chi_\mathrm{p}$ by setting $\chi_1^\perp=0.99$. We adjust the source distance to ensure a comparable SNR. In this case, the relationship between the estimated minimum measurable $e_{22,10}$ and total mass is less straightforward, but the minimum measurable $e_{22,10}$ is always lower than it is for its non-spinning counterpart. 

The difference can be understood from the fact that, in GWs from spin-precessing binaries, the envelope of the spin-precession-induced amplitude oscillations is also altered by the presence of eccentricity. As the mass ratio $q$ gets more unequal for a fixed mass, the number of these precession cycles in-band increases, leading to improved eccentricity measurability. We illustrate this effect in Figure 6 in the Appendix.
We include in Figure \ref{fig:scatter-plot-pyefpe} the estimated minimum measurable eccentricity for maximally spin-precessing injections with $M=45$~M$_\odot$ and varying $q$, illustrating that more unequal mass ratios improve measurability in this case.
For fixed $m_2=1.5$~M$_\odot$ injections with $\chi_\mathrm{p} =\mathrm{max.}$, we interpret the turn-over at $M\approx10$~M$_\odot$ as the point at which the shortening effects of increasing $M$ on the waveform are counteracted by the increasing number of precession cycles in-band.

\subsection{Astrophysical implications of detecting eccentric NSBHs}

\begin{figure*}
    \centering
    \includegraphics[width=\linewidth]{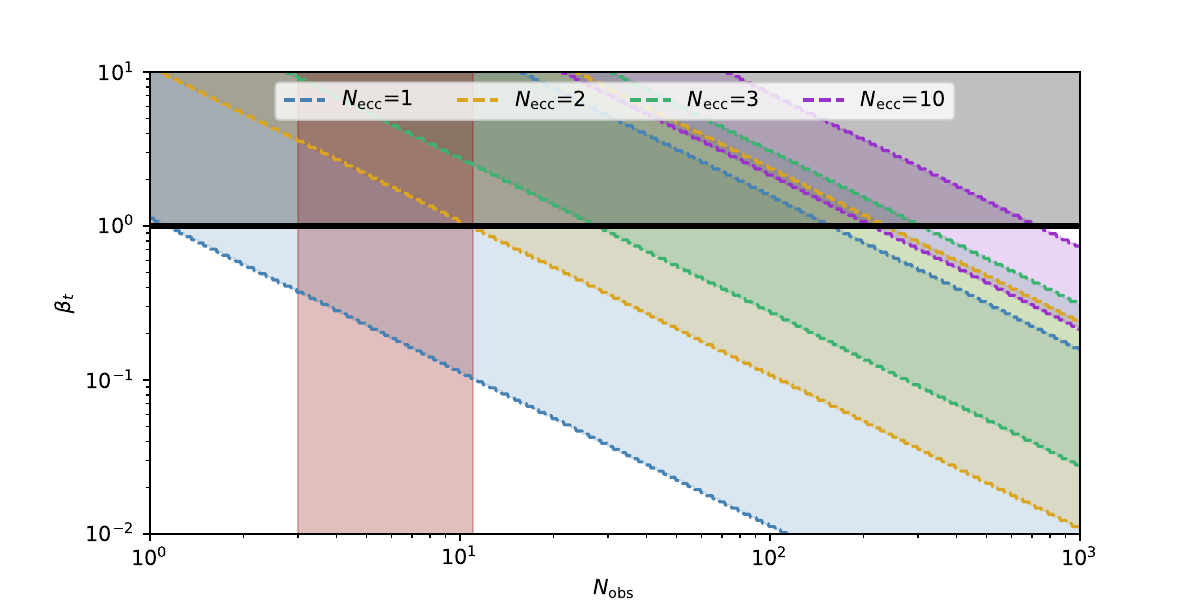}
    \caption{Constraints on the detectable branching fraction of NSBH-producing field triples, $\beta_t$, as a function of the number of NSBH observations, $N_\mathrm{obs}$, given a number of eccentric NSBH observations $N_\mathrm{ecc}$. Diagonal shaded coloured bands encompass the 95\% symmetric credible region of the likelihood on $\beta_t$. The limit $\beta_t=1$ is marked with a thick horizontal black line. Observations above this limit would indicate either problems with the astrophysical simulations, or more eccentric observations than can be explained by this formation channel alone. The vertical red shaded band encompasses the range of possible $N_\mathrm{obs}$ so far, including marginal candidates and candidates with ambiguous secondary masses to reach the maximum.}
    \label{fig:2D_probs}
\end{figure*}

\begin{figure}
    \centering
    \includegraphics[width=\linewidth]{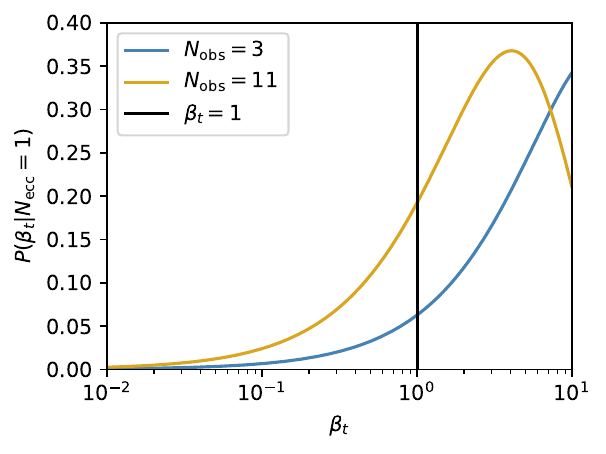}
    \caption{The probability distribution over the triple branching fraction $\beta_t$ given number of NSBH observations $N_\mathrm{obs}=3$ (blue) and $N_\mathrm{obs}=11$ (gold) assuming $N_\mathrm{ecc}=1$. }
    \label{fig:1D_probs}
\end{figure}

In the left-hand panel of Figure \ref{fig:e_10_recovered_pyefpe}, we compare the histograms of the simulation-output source-frame eccentricities with a histogram of the source eccentricities of injections found using our quasi-circular aligned-spin mock search, including the effects of redshifting and marginalising over variations in extrinsic parameters. Our mock search recovers \new{$3.9\%$} of the total simulated population. If we compute the mock search using eccentric and spin-precessing templates, we recover \new{$8.5\%$} of our total simulated population. Alternatively, if we alter our simulated population to have only aligned spins and zero eccentricity, the aligned-spin quasi-circular mock search recovers \new{$10.5\%$} of systems. The relative detection efficiency of the search is therefore \new{$46\%$} compared to a search containing the same physics as the injected population, and \new{$37\%$} compared to the recovery of an equivalent non-eccentric non-spin-precessing population.

We find that the search sensitivity drops steeply above \new{$e_{22,10}=0.05$}, and that in the region of the eccentricity posterior for GW200105, there are about \new{eight} times as many \new{signals that are found in the eccentric and spin-precessing search as found in the quasi-circular search}. In the right-hand panel of Figure \ref{fig:e_10_recovered_pyefpe}, we show the percentage of signals that are found with an eccentricity greater than or equal to $e_{22,10}$. When $e_{22,10}\geq0.01$, only $2\%$ of sources are recovered.

In \citet{Zevin2021}, $e_{10}=0.05$ was used as the minimum detectable eccentricity threshold for binary BH mergers, based on the recovery of GW150914-like injections using Bayesian inference \citep{Lower2018}. Using the optimistic overlap method, we estimate that NSBH-like systems have similar detectable eccentricities of $e_{22,10} \approx \mathcal{O}(0.01)$. In fact, the overlap estimates of \citet{Lower2018} for GW150914-like systems are slightly lower than our estimates for NSBHs, possibly due to using a different waveform model or the higher SNR of the GW150914-like injections.

Fortunately, we do not require an accurate and universal ``minimum detectable eccentricity'' to establish the astrophysical implications of detecting eccentric NSBHs; we merely require an arbitrary eccentricity threshold above which we are confident that eccentricity measurement would be achieved. In accordance with standard practice for binary BH eccentricity measurability, we choose $e_{22,10}=0.05$ \citep{Lower2018, Zevin2021}. In the simulations of \citet{StegmannKlencki2025}, most of the NSBH mergers have eccentricities above this threshold: $\approx80\%$ of NSBHs have $e_{22,10} \gtrsim 0.05$. Due to the low detectability of highly-eccentric inspiral-dominated systems, though, only \new{$2.8\%$} of these are detected in our mock search. We therefore set the ``above threshold'' eccentric fraction, i.e. the fraction of systems that we are confident are detectable and measurably eccentric, to \new{$\xi_\mathrm{thresh} = 0.80 \times 0.028 = 0.0224$}. Using the same approach and notation as \citet{Zevin2021}, in Figure \ref{fig:2D_probs}, we plot the symmetric $95\%$ credible intervals around $\beta_t$, the branching fraction of field triples to the total population of observed NSBHs. If the number of observed NSBHs is $N_\mathrm{obs}=3$, and if GW200105 is truly eccentric so $N_\mathrm{ecc}=1$, then field triples may constitute \new{$\geq40\%$} of observed NSBHs.

There are likely more than three NSBHs in the catalogue so far. In addition to GW190814, which may be an NSBH with an unusually massive NS secondary, GWTC-2 included the low-significance NSBH candidate GW190426\_152155 \citep{GWTC-2}. In GWTC-2.1, one confident candidate, GW190917\_114630, has a secondary consistent with being an NS ($m_2\approx2.1$~M$_\odot$), although it is identified by the search pipeline as a binary BH; there are two further low-significance NSBH candidates, GW190531\_023648  and GW190917\_114630 \citep{GWTC-2-1}. In GWTC-3, in addition to GW200115 and GW200105, two potential NSBHs---GW191219\_163120 and GW200210\_092254---were reported, although the former has an uncertain astrophysical probability, and the latter, like GW190814, has an ambiguous secondary mass \citep{GWTC-3}. In GWTC-4, both GW230529 and confident NSBH candidate GW230518\_125908 were reported \citep{GWTC-4}; however, the latter was not included in downstream analyses (e.g., the rates and populations analysis \citep{GWTC-4-astro-pop}) due to being detected outside of official observing, during the engineering run. 

We include in Figure \ref{fig:2D_probs} a vertical red band spanning $3 \leq N_\mathrm{obs} \leq 11$, demonstrating the uncertainty on the number of NSBH detections. With $N_\mathrm{obs}=11$ and $N_\mathrm{ecc}=1$, the lower $90\%$ credible limit on $\beta_t$ is pushed down to \new{$\approx10\%$}, while the detections remain consistent with $100\%$ of observed NSBHs originating in field triples at $90\%$ credibility. If no more eccentric NSBHs are detected, the upper $90\%$ credible limit on $\beta_t$ can begin to be constrained to $<1$ \new{when we reach $N_\mathrm{obs} = 157$}.

In Figure \ref{fig:1D_probs}, we plot the probability distribution over $\beta_t$ assuming $N_\mathrm{ecc}=1$ for $N_\mathrm{obs}=3$ and $N_\mathrm{obs}=11$. \new{In both cases}, the peak and majority of the likelihood distribution is in the unphysical region above $\beta_t=1$, implying \new{an overabundance} of eccentric mergers given the expected detection rates. 

\section{Discussion \& Conclusion}
\label{sec:discussion}

In this work we investigate the measurability and detectability of eccentricity in GW signals from merging NSBHs. We find that, while NSBHs similar to those that have been detected so far have similar minimum measurable eccentricities to binary BHs of $\mathcal{O}(0.01)$, those that exhibit significant higher-mode content may have measurable eccentricities an order of magnitude lower. We also find that the presence of non-zero precessing spin may decrease the minimum measurable eccentricity (i.e., improving our ability to measure eccentricity) for more unequal-mass NSBH mergers. By simulating a quasi-circular templated search for a realistic population of NSBH mergers produced by field triples, we find that many signals from these binaries would not be found in current detectors and pipelines, with \new{$96.1\%$} having an SNR computed against a quasi-circular, aligned-spin template that is below our assumed detection threshold. 

This low recovery fraction has two implications: firstly, that a large population of triple-origin NSBHs could be merging in the Universe without being found by our searches; and secondly, that a single detection of an eccentric NSBH implies a high rate of NSBH mergers from triples. Indeed, we find that if we have observed only three NSBH mergers and one of them is measurably eccentric, then the likelihood for the branching fraction $\beta_t$ of NSBHs from triples peaks above the physical upper prior limit of $\beta_t=1$. \new{Even if all 11 marginal and ambiguous NSBH candidates in the GW transient catalogue so far are counted as detections, then one measurably eccentric merger yields a $\beta_t$ that is still above $1$.} 

If triples are efficiently producing NSBH mergers, then they may also efficiently produce binary BH mergers. We speculate that the fraction of detectably eccentric mergers should be smaller for a population of binary BHs evolved with the same assumptions as \citet{StegmannKlencki2025}, because the fractional mass loss during compact-object formation and compact-object natal kicks are smaller for BHs than NSs, so more distant tertiaries can remain bound and lead to less aggressive orbital driving. This is consistent with predictions of eccentricity distributions in binary BHs from triples \citep[e.g.,][]{Antonini2016, Antonini2017, Liu2018, Dorozsmai2025}; however, we note that the results of these triple-evolving binary BH simulations cannot easily be compared to the results of \citet{StegmannKlencki2025}, since they employ different assumptions and cover different regions of parameter space. The relative rates of binary BH and NSBH mergers from triples are also uncertain, given the lack of self-consistent simulations of triples producing both types of binary merger. For binary NSs, \citet{HamersThompson:2019:triples} find that merger rates can be an order of magnitude higher than for NSBHs if natal kicks are low, but can be lower if natal kicks are high. A full population synthesis study with consistent assumptions, aiming to quantify the relationship between the NSBH merger rate, the binary BH merger rate, and the binary NS merger rate from triples, is left for future work.

In the simulations of \citet{StegmannKlencki2025}, for every triple-origin NSBH produced with $e_{22,10} < 0.05$, there are $\approx4$ produced with higher eccentricity, which are less likely to be detected. As shown in the right-hand panel of Figure \ref{fig:e_10_recovered_pyefpe}, our mock search suggests that \new{$<1\%$} of NSBHs with an eccentricity of $e_{22,10}\approx0.3$---similar to that claimed for GW200105---would be detected with a quasi-circular aligned-spin search. Consequently, for every eccentric NSBH detected with a quasi-circular aligned-spin search, there could be \textbf{many} more signals with similar or higher eccentricities that are missed. This motivates the development of search strategies for inspiral-dominated signals with a wide range of total masses, eccentricities, and spin orientations.

\section*{Acknowledgements}

\new{We sincerely thank Alex Nitz for thoroughly assessing our work, and in particular for spotting an error in our computations that led to artificially high recovered detection fractions from our searches in v1 of this manuscript. We also} thank Nick Loutrel, Antoni Ramos-Buades, P.J. Nee, Nihar Gupte, and Aldo Gamboa for helpful discussions, and thank Tom Dent, Ben Patterson, Alejandra Gonzalez, and Arif Shaikh for comments on the draft.
I.M.R-S acknowledges support from the Science and Technology Facilities Council Ernest Rutherford Fellowship grant number UKRI2423.
M.Z. gratefully acknowledges funding from the Brinson Foundation in support of astrophysics research at the Adler Planetarium. 
The authors are grateful for computational resources provided by the LIGO Laboratory and supported by National Science Foundation Grants PHY-0757058 and PHY-0823459.

\section*{Data availability statement}

The data underlying the analyses in this paper will be made available on request to the corresponding author.

\bibliographystyle{mnras}
\bibliography{ecc-nsbh}

\appendix

\section{Effect of sampling frequency and inclination on SEOBNRv5EHM overlaps}

\begin{figure*}
    \centering
    \includegraphics[width=\linewidth]{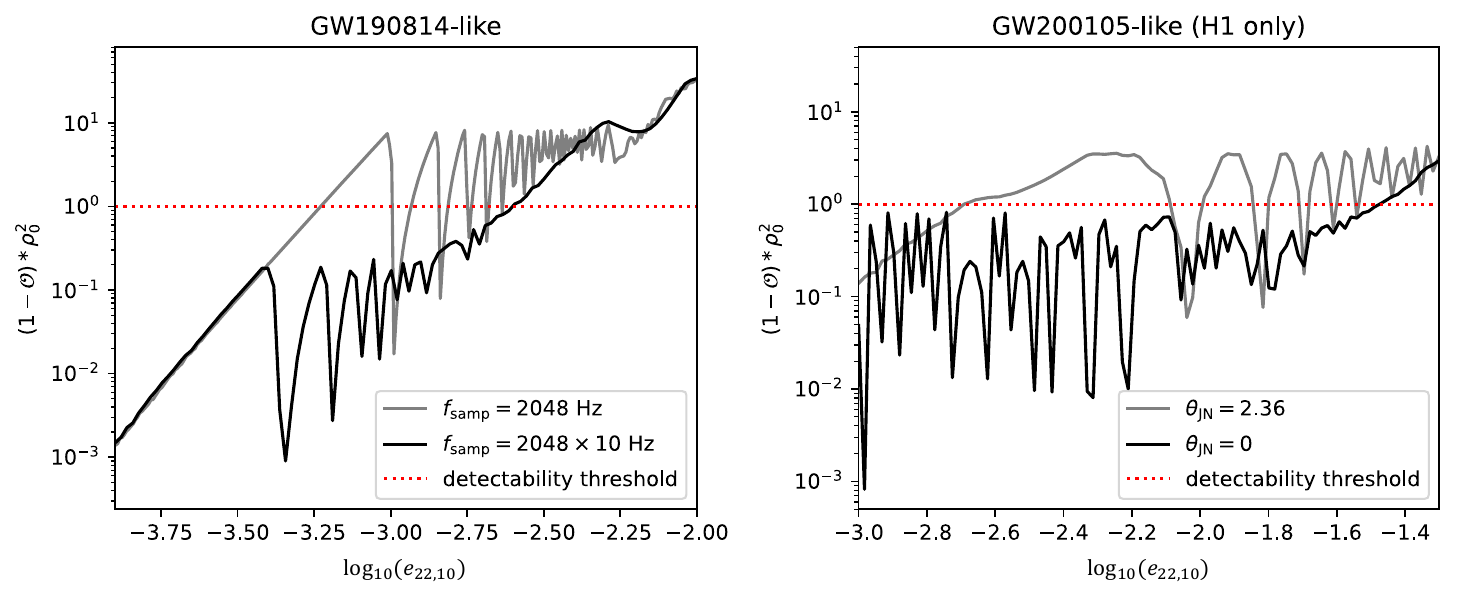}~
    \caption{Oscillations seen in the overlap calculated using \textsc{SEOBNRv5EHM} for (left) a GW190814-like injection with different sampling frequencies, and (right) a GW200105-like injection with different source inclinations. All other parameters are as shown in Table \ref{tab:settings}. We see that the highest value of $e_{22,10}$ at which the metric $(1 - \mathcal{O}) * \rho^2_0$ becomes $> 1$ is similar in all variations.}
    \label{fig:compare-oscillations}
\end{figure*}

In Figure \ref{fig:compare-oscillations}, we demonstrate the influence of varying the sampling frequency $f_\mathrm{samp}$ and inclination angle $\theta_\mathrm{JN}$ on the overlap calculated with the \texttt{SEOBNRv5EHM} waveform model. We believe these differences arise because we marginalise over a uniform distribution of input relativistic anomalies, which does not remain uniform as the binary evolves, leading to some orientations being more well-sampled than others at merge. The fact that increasing the sampling frequency reduces the scale of these oscillations in the phase-time maximised overlap implies that a more uniform distribution of input relativistic anomaly is achieved when sampling frequency is higher. When the system has $\theta_\mathrm{JN}=0$, the amplitude of the oscillations again reduces. This is likely related to the fact that the tilt of the orbit changes the impact of the relativistic anomaly parameter on the detected waveform. 

\section{Effect of spin-precession on eccentricity measurability}

\begin{figure*}
    \centering
    \includegraphics[width=\linewidth]{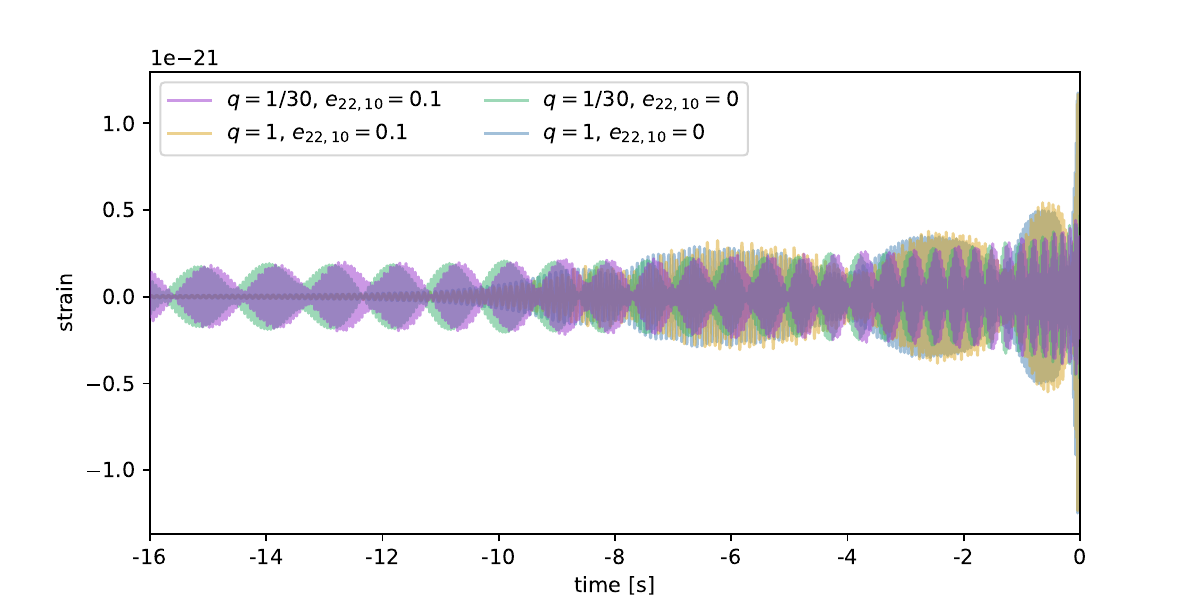}
    \caption{Comparison of maximally spin-precessing \texttt{pyEFPE} waveforms with different mass ratios and eccentricities. All have $\chi_1^\perp=0.99$ and $M=45$~M$_\odot$.}
    \label{fig:waveforms}
\end{figure*}

In Figure \ref{fig:waveforms}, we demonstrate the influence of mass ratio $q$ on the number of spin-precession cycles in-band. When $q=1$ the cycles last longer and their cadence is lower than when $q$ is more unequal. This leads to an improvement in eccentricity measurability for unequal-mass, spin-precessing binaries, since eccentricity alters the envelope of these precession cycles.

\bsp	
\label{lastpage}
\end{document}